\documentclass[final,5p,times,twocolumn]{elsarticle}
\usepackage{amssymb}
\journal{Physics Letters B}

\usepackage{mathtools}
\usepackage{epsfig}
\newcommand{\beqn}{\begin{eqnarray}}
\newcommand{\eeqn}{\end{eqnarray}}
\newcommand{\eq}[1]{(\ref{#1})}

\newcommand{\cO}{{\cal O}}

\newcommand{\cL}{{\cal L}}

\newcommand{\vort}{{\mathrm{vort}}}

\newcommand{\Z}{{\mathbb Z}}
\newcommand{\bs}{\boldsymbol}

\newcommand{\avr}[1]{{\langle #1 \rangle}}
\newcommand{\aavr}[1]{\avr{\!\avr{ #1 }\!}}

\begin{document}

\begin{frontmatter}

\title{Conformal magnetic effect at the edge: a numerical study in scalar QED}

\author[IDP,FEFU]{M. N. Chernodub}
\author[FEFU]{V. A. Goy}
\author[FEFU]{A. V. Molochkov}
\address[IDP]{Institut Denis Poisson UMR 7013, Universit\'e de Tours, 37200 France}
\address[FEFU]{Laboratory of Physics of Living Matter, Far Eastern Federal University, Sukhanova 8, Vladivostok, 690950, Russia}

\begin{abstract}
Quantum polarization effects associated with the conformal anomaly in a static magnetic field background may generate a transverse electric current in the vacuum. The current may be produced either in an unbounded curved spacetime or in a flat spacetime in a physically bounded system. In both cases, the magnitude of the electric current is proportional to the beta-function associated with renormalization of the electric charge. In our article, we investigate the electric current density induced by the magnetic field in the vicinity of a Dirichlet boundary in the scalar QED. Using first-principle lattice simulations we show that the electric current, generated by this ``conformal magnetic effect at the edge'' (CMEE), is well described by the conformal anomaly provided the conformal symmetry is classically unbroken. Outside of the conformal limit, the current density is characterized by an anomalous power law near the edge of the system and by an exponential suppression of the current far away from the edge.
\end{abstract}


\end{frontmatter}

\section{Introduction}

Quantum anomalies lead to a large variety of unusual transport phenomena such as the chiral magnetic effect~\cite{Fukushima:2008xe}, the chiral vortical effect~\cite{Son:2004tq} and their generalizations~\cite{Landsteiner:2012kd}. The anomalies generate vector and axial currents of chiral fermions in a background of electromagnetic field and in a rotating environment, at finite and zero temperatures, in dense and/or chirally imbalanced media. Experimental signatures consistent with some of these transport phenomena were observed in diverse systems, from quark-gluon plasma created in ultrarelativistic heavy-ion collisions to crystals of Dirac and Weyl semimetals~\cite{Kharzeev:2013ffa}.

The chiral magnetic and chiral vortical effects are associated with the breaking of axial and mixed axial-gravitational symmetries of corresponding classical systems. Anomalous breaking of another classical symmetry, the conformal invariance, may also lead to a particular form of transport phenomena such as the scale magnetic and scale electric effects in a curved gravitational background~\cite{Chernodub:2016lbo}. In was recently proposed that the scale magnetic effect may be realized in Dirac and Weyl semimetals as a Nernst effect, i.e. the generation of an anomalous electric current normal to a temperature gradient that drives the system slightly out of equilibrium~\cite{Chernodub:2017jcp}. Technically, the scale electromagnetic effects appear due to the anomalous gauge-gauge-graviton $TJJ$ vertex which contributes in anomalous conformal action and involves the correlator of the energy-momentum tensor and two vector currents~\cite{ref:TJJ}.

In the presence of the background magnetic field the conformal anomaly may also generate an electric current in spatially bounded systems~\cite{McAvity:1990we,Chu:2018ksb}. The anomalous current is normal to the axis of the magnetic field and is tangential to the edge of the system. It arises in a vacuum in the absence of the background electric field with zero chemical potentials and at zero temperature. Similarly to the case of the scale electromagnetic effects -- mediated by the conformal anomaly in a curved spacetime -- the anomalous edge current is proportional to the beta function associated with the renormalization of the electric charge. We refer to the current generated at the boundary as the Conformal magnetic effect at the edge or as the Conformal magnetic edge effect (CMEE).

Contrary to the axial and mixes axial-gravitational anomalies, the conformal anomaly may emerge both in fermionic and in bosonic systems. Therefore we expect that the CMEE may appear in both these cases. 

In our paper we numerically investigate, from the first principles, the generation of the boundary electric current in the lattice formulation of the (3+1) dimensional Abelian Higgs model (AHM) following the analytical studies of Refs.~\cite{McAvity:1990we,Chu:2018ksb}. The AHM possesses two phases: the phase with a spontaneously broken $U(1)$ symmetry (the condensed, or superconducting phase) and an unbroken (Coulomb) phase with a massless scalar field. The model in the unbroken phase is often called as ``the scalar QED''. Despite the fact that the broken and unbroken phases are analytically connected, in a certain region of the coupling space these phases are separated by a first-order phase transition which ends in an end-point where the transition becomes of the second order.~\cite{ref:AHM}. We are interested in a region in a vicinity of the second-order phase transition where the mass gap vanishes and the model approaches the conformal limit. 

The structure of the paper is as follows. In Section~\ref{sec:theory} we briefly discuss a qualitative physical picture behind the CMEE in the conformal theory. Section~\ref{sec:model} is devoted to the details of our lattice model. In Section~\ref{sec:conformal} we search for the conformal point and give details of the numerical simulations. We describe our numerical results for the local density of the electric current in Section~\ref{sec:current}. Our conclusions are summarized in the last section.

\section{Conformal magnetic effect at the edge}
\label{sec:theory}

In Refs.~\cite{McAvity:1990we,Chu:2018ksb} it was shown that for a general class of spatially bounded quantum field theories with $U(1)$ gauge symmetry, the conformal anomaly generates an electric current in the vicinity of the boundary. The current takes the following form:
\beqn
{\bs j}({\bs x}) = - f({\bs x}) \, {\bs n} \times {\bs B}
\label{eq:j:conformal}
\eeqn
where ${\bs B}$ is the magnetic field and ${\bs n}$ is a normal vector pointing outwards to the boundary of the surface. The anomalous factor
\beqn
f({\bs x}) = \frac{2 \beta(e)}{e^2} \frac{1}{x_\perp}\,,
\label{eq:nu}
\eeqn
is a function of the tangential distance~$x_\perp$ from the boundary to the point ${\bs x} = ({x_\perp,{\bs x}_\|})$ in the bulk where the current is generated. In our notations $x_\perp = x$ and ${\bs x}_\| = (y,z)$, and $e = |e|$ is the elementary electric charge.

Physically, the electric current~\eq{eq:j:conformal} appears due to quantum creation of particle-antiparticle pairs. In the presence of the background magnetic field the particles and antiparticles will propagate along a closed cyclotron orbit and eventually annihilate each other. Near a reflective boundary of the system, this circular cyclotron motion is not possible due to collisions of the created (anti)particles with the boundary. The created particles and antiparticles will thus propagate in ``skipping orbits'' in opposite directions along the edges of the system, thus creating a double electric current tangential to the boundary.

The anomalous coefficient~\eq{eq:nu} depends on the beta function $\beta = \beta(e)$ associated with the renormalization of the electric charge~$e$. The induced electric current~\eq{eq:j:conformal} is tangential to the boundary of the system and is, simultaneously, normal to the axis of magnetic field. The electric current is generated by the magnetic field along the edge of the zero-temperature system in the absence of matter and electric field. The electric current at the boundary~\eq{eq:j:conformal} is proportional to the beta function associated with the renormalization of electric charge via Eq.~\eq{eq:nu}, thus highlighting the conformal--anomalous nature of this Conformal magnetic edge effect (CMEE).

In our paper we concentrate on scalar QED (sQED) which is much easier to simulate numerically compared to the usual QED with light fermion(s). The one-loop $\beta$ function of sQED with one charged bosonic species is as follows:
\beqn
\beta^{{\text{1-loop}}}_{{\text{sQED}}} = \frac{e^3}{48 \pi^2}\,,
\label{eq:beta:sQED}
\eeqn
and therefore the anomalous coefficient~\eq{eq:nu} is given by the simple expression:
\beqn
f_{cQED} (x_\perp) = \frac{1}{24 \pi^2 x_\perp}\,.
\label{eq:nu:cQED}
\eeqn

The electric current~\eq{eq:j:conformal} is tangential with respect to the surface and normal to to the external magnetic field ${\bs B}$. In the conformal limit the tangential component of the total electric current induced by the CMEE~\eq{eq:j:conformal},
\beqn
J^{\mathrm{tot}}_\| = \int_0^{\infty} j_\|(x_\perp)  \, d x_\perp\,,
\label{eq:int:j}
\eeqn
is a linear function of the background magnetic field:
\beqn
J^{\mathrm{tot}}_\| = g eB\,,
\eeqn
where the proportionality coefficient $g$ diverges both in infrared and ultraviolet limits:
\beqn
g  = \frac{1}{24 \pi^2} \ln \frac{\lambda_{IR}}{\lambda_{UV}}\,.
\label{eq:gamma:conf:th}
\eeqn

In the conformal limit the infrared cutoff~$\lambda_{IR}$ in Eq.~\eq{eq:gamma:conf:th} should be of the order of the double radius of the circular trajectories of the charged particles in the lowest Landau level, $\lambda_{IR} \sim 2 R_{LLL} = 2/\sqrt{|eB|}$. Indeed the particles which are localized deeper in the bulk move along closed circular trajectories and thus they do not contribute to the boundary current. The external classical magnetic field breaks explicitly the conformal symmetry of the problem, so that the appearance of this fact is well justified.

The ultraviolet cutoff $\lambda_{UV}$ in Eq.~\eq{eq:gamma:conf:th} should be determined by a typical smallest scale of the material at which the continuum description of particle's motion is no more applicable (the value of the cutoff $\lambda_{UV}$ could be of the order of an interatomic distance, for example), and the conformal symmetry is again broken explicitly. In our simulations the ultraviolet scale is naturally given by the lattice spacing, $\lambda_{UV} = a$.

We will show that outside the conformal window the total current~\eq{eq:int:j} is a finite quantity which needs no regularization. 

\section{Model}
\label{sec:model}

We numerically investigate the CMEE~\eq{eq:j:conformal} in the scalar QED with the conformally invariant Lagrangian
\beqn
\cL_{\text{sQED}} = - \frac{1}{4} F_{\mu\nu} F^{\mu\nu}
+  \left[\left( \partial_\mu - i e A_\mu \right) \phi \right]^* \left( \partial^\mu - i e A^\mu \right) \phi \,,
\label{eq:sQED:1}
\eeqn
where $\phi$ is the charged Higgs field and $A_\mu$ is the gauge field with the field strength $F_{\mu\nu} = \partial_\mu A_\nu - \partial_\nu A_\mu$. In order to numerically simulate the theory~\eq{eq:sQED:1} we employ the lattice Abelian Higgs model (AHM) with a certain potential on the scalar field, and then search for the conformal point where both scalar and gauge fields become (almost) massless as in Eq.~\eq{eq:sQED:1}.

The Lagrangian of the AHM in the background magnetic field is given by the following formula:
\beqn
& & 
\hskip -10mm 
S = \beta_{\mathrm{latt}}\sum_x \sum_{\mu<\nu = 1}^4 \left( 1 - \cos \theta_{x,\mu\nu} \right)
 \label{eq:S:AHM} \\
& & 
\hskip -7mm 
+ \sum_{x} \sum_{\mu=1}^4 \left| \phi_x -  e^{i (\theta_{x\mu} + \theta^B_{x\mu})} \phi_{x+\hat\mu} \right|^2
+ \sum_{x} \left( - \kappa  \left| \phi_x \right|^2 + \lambda \left| \phi_x \right|^4 \right) \,,
\nonumber 
\eeqn
where the complex scalar field $\phi_x$ and the vector gauge field $\theta_{x\mu}$ are the dynamical fields while the vector field $\theta^B_{x\mu}$ represents a background (classical) magnetic field. The lattice gauge coupling $\beta_{\mathrm{latt}}$ is related to the electric charge $e$ as $\beta_{\mathrm{latt}} = 1/e^2$.\footnote{We introduced the subscript ``latt'' in order to distinguish the lattice coupling $\beta_{\mathrm{latt}}$ in the action~\eq{eq:S:AHM} from the $\beta$ function in the anomalous coefficient~\eq{eq:nu}.} The bare couplings $\kappa$ and $\lambda$ are associated, respectively, with quadratic and quartic terms of the scalar field $\phi$. 

Action~\eq{eq:S:AHM} is invariant under the Abelian gauge transformations: $\theta_{x\mu} \to \theta_{x\mu} + \omega_x - \omega_{x+\hat\mu}$, $\phi_x \to e^{i \omega_x} \phi_x$, where $\omega_x$ is an arbitrary real scalar function defined at the sites of the lattice. The background gauge field is not transformed under the gauge transformations: $\theta^B_{x\mu} \to \theta^B_{x\mu}$.

The AHM model~\eq{eq:S:AHM} is also characterized by the physical lattice spacing $a$ given by the physical length of an elementary lattice link. In the naive continuum limit $a \to 0$ the dimensionless fields $\theta_l$ and $\phi$ are related to their continuum counterparts (both of the dimension of mass) via the relations $A_\mu(x) = \theta_{x,\mu}/a$ and $\varphi(x) = \phi_x/a$. The value of the lattice spacing $a$ is usually determined by matching  dimensionless lattice results to known dimensionful experimental data (for example, to a known mass of an excitation). In our paper we work in dimensionless lattice units suitable for the conformal limit of the model.

In the lattice model~\eq{eq:S:AHM} the gauge field $\theta_l$ is a compact variable because the action is invariant under the shifts $\theta_l \to \theta_l + 2 \pi n_l$ where $n_l \in \Z$. The compactness of the gauge field implies the existence of Abelian monopoles in the lattice theory, and, consequently, the presence of a confining phase in the strong coupling region with $e \gtrsim 1$ (at small lattice coupling $\beta_{\mathrm{latt}}$). In our numerical calculations we keep the lattice gauge coupling sufficiently large so that the monopoles are suppressed. We, consequently, work far away from the confining phase so that the compact nature of the lattice gauge field $\theta_l$ does not influence our results.

We study the CMEE in a zero-temperature model at the lattice $N_1 \times N_s^3$ with periodic boundary conditions in all four directions. The take $N_1 > N_s$ so that the long direction $N_1$ corresponds to a spatial dimension. We impose the reflective Dirichlet boundary in the spatial plane located at the middle of the lattice: $x_1 = N_1/2$. It is convenient to introduce the coordinate $x_\perp = x_1 - N_1/2$ so that the position of the boundary corresponds to $x_\perp = 0$:
\beqn
\phi_x {\biggl|}_{x_\perp = 0} = 0\,.
\label{eq:Dirichlet:boundary}
\eeqn
In our simulations we choose $N_1 = 48$ in the spatial direction perpendicular to the Dirichlet boundary and $N_2=N_3=N_4 \equiv N_s = 32$ in all other directions.

In order to study the anomalous current generation we introduce a static uniform magnetic-field background
\beqn
\theta^B_{x,12} = \frac{2 \pi k}{N_1 N_2}\,,
\label{eq:eB:k}
\eeqn
along the $x_3$ axis which is parallel to the Dirichlet surface~\eq{eq:Dirichlet:boundary}. The corresponding gauge field is parameterized as follows:
\beqn
& & \theta^B_{x,2} = \frac{2 \pi k}{N_1 N_2} x_1\,, \qquad\ \ k = 0, 1, \dots , \frac{N_1 N_2}{2}\,,
\label{eq:theta:B} \\
& & 
\theta_{x,1}^B{\biggl|}_{x_1=N_1-1} = -\frac{2\pi k}{N_2} x_2, 
\qquad 
\theta^B_{x,3} =  \theta^B_{x,4} = 0\,,
\eeqn
while other components of the gauge field are zero.

In Eq.~\eq{eq:theta:B} the integer number $k$ characterizes the strength of the magnetic field. The quantization of the electromagnetic gauge field~\eq{eq:theta:B} is appears as a result of the periodic spatial boundary conditions $e^{i \theta^B_{x+L,\mu}} = e^{i \theta^B_{x,\mu}}$, while the bound on $k$ from above stems from the compactness of the gauge field~$\theta_l$. The value $k$ is equal to the number of elementary fluxes introduced on the lattice in the $(x_1,x_2)$ plane. The maximal value of $k$ in Eq.~\eq{eq:theta:B} corresponds to the lattice half-filled with Abrikosov vortices so that the physical vortex density is given by the lattice ultraviolet cutoff $\rho_\vort \sim a^{-2}$. We will work at weak magnetic fields $\theta^B$ which are sufficiently far away from the artificially large values of the integer number $k$. The lattice magnetic field~\eq{eq:eB:k} is related to the magnetic field $B$ in the continuum limit via the relation $\theta^B_{x,12} = eB a^2$. A direct calculation at the conformal point is rather challenging due to large correlation lengths which require time-consuming simulations at large volumes.

\section{Conformal point}
\label{sec:conformal}

In this article we are interested in the results close to a conformal region of the parameter space for which the mass gap vanishes. In the conformal region the longest correlation length(s) become(s) infinite thus signaling the presence of a second-order phase transition. Lattice simulations at a very point of a second-order transition are impossible form a practical point of view since at the phase transition the field fluctuations are very large being at the same time sensitive to the volume of the system (in other words, the dynamics of the fields in the bulk of the system are affected by the boundaries of the system due to the divergent correlation length). Therefore in our simulations use the following strategy: (i) we consider a path in the parameter space which approaches a point near the second-order phase transition; (ii) we make calculations at a sufficiently large set of points at the chosen path and then extrapolate the results to the nearly-conformal point. 

In this article we simulated the model~\eq{eq:S:AHM} at the fixed parameters $\beta_{\mathrm{latt}} = 4$ and $\lambda = 10$. We generated field configurations using of a Hybrid Monte Carlo algorithm~\cite{ref:Gattringer,ref:Omelyan} and performed simulations on Nvidia GPU cards. To achieve acceptable statistics we used from $10^6$ to $33\times 10^6$ trajectories per each value of the background magnetic field. In order to accelerate calculations and reduce write operations we accumulated mean values per each 100 trajectories only as even in this case our simulations generated about 1.5TB of the data [for example, each configuration of electric current $J=J(x_1,x_2,\mu)$ in the $(x_1,x_2)$ plane gives us $2^{12}$ double variables (32KB) per one trajectory for the $32^4$ lattice]. We used binning for correct error estimations for our observables.

\begin{figure}[!thb]
\begin{center}
\includegraphics[scale=0.5,clip=true]{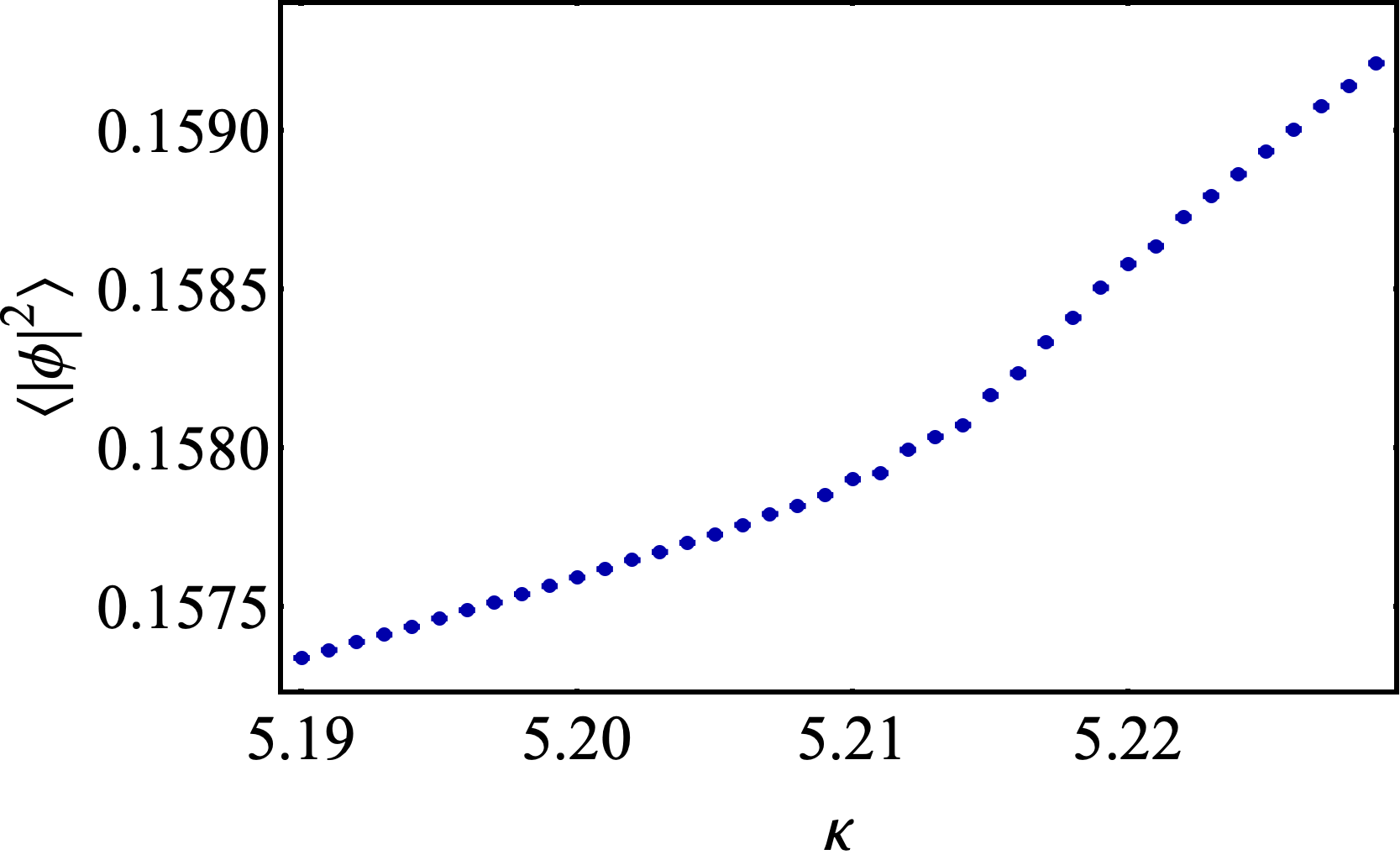}\\[0mm] 
\hskip 15mm (a)
\\[4mm]
\hskip 8mm \includegraphics[scale=0.44,clip=true]{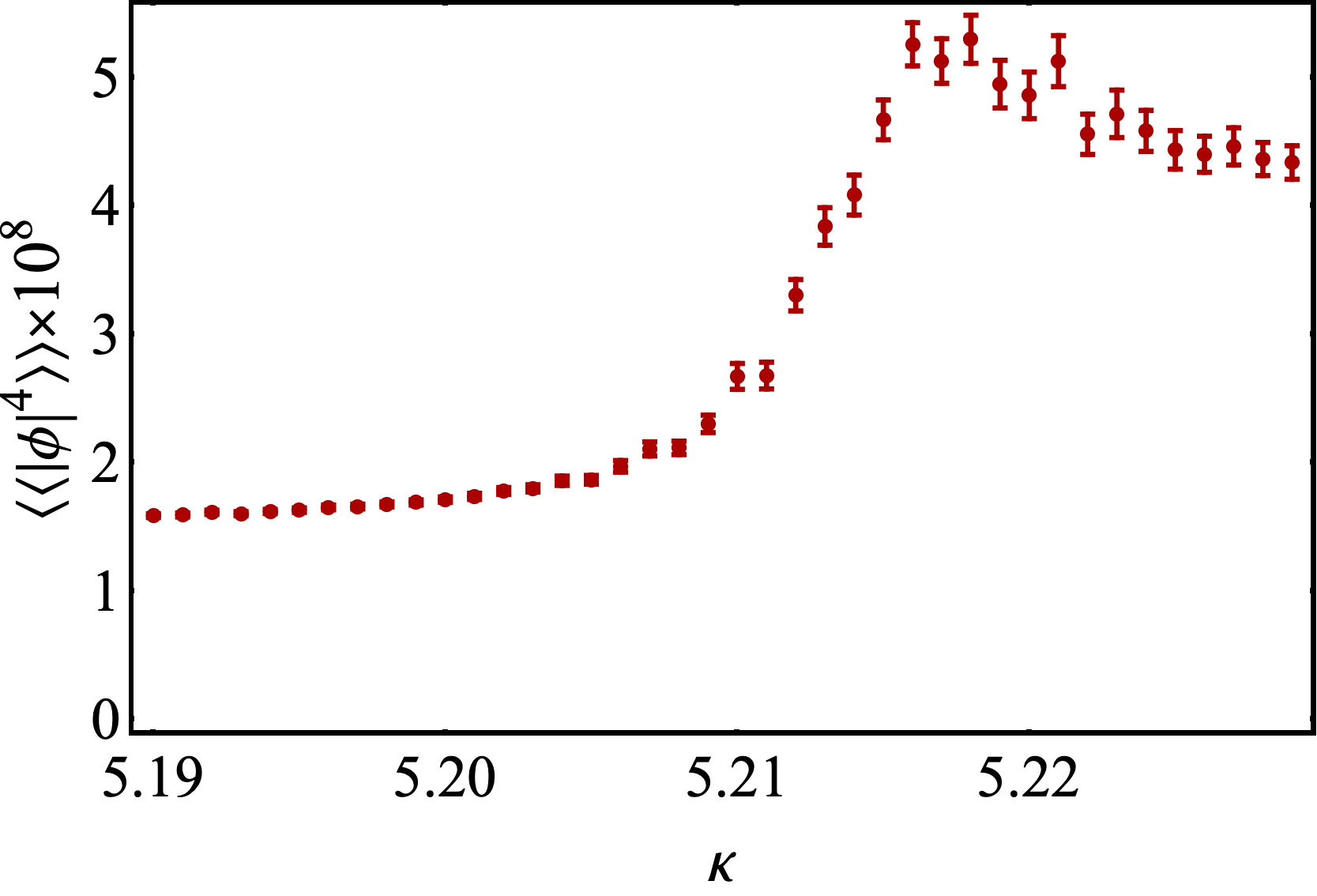}\\[0mm] 
\hskip 15mm (b)
\end{center}
\vskip -3mm
\caption{The expectation value of the scalar field squared $\avr{|\phi|^2}$ (the upper plot) and the susceptibility $\aavr{|\phi|^4}$ (the lower plot) as functions of the hopping parameter $\kappa$ at the fixed gauge coupling $\beta_{\mathrm{latt}} = 4$ and the quartic coupling $\lambda {=} 10$.}
\label{fig:conformal:point}
\end{figure}

By varying the hopping parameter $\kappa$ we find a continuous transition that occurs at around the point $\kappa \simeq 5.215$. At this point the expectation value of the squared scalar field $\avr{|\phi|^2}$ exhibits a knee, Fig.~\ref{fig:conformal:point}(a), while the corresponding susceptibility, $\aavr{|\phi|^4} \equiv \avr{|\phi|^4} - \avr{|\phi|^2}^2$ shows a wide smooth maximum, Fig.~\ref{fig:conformal:point}(b). The broken and unbroken are located at right and left sides of the transition, respectively. We do not need to perform a renormalization procedure in our studies, so that the unrenormalized expectation value $\avr{|\phi|^2}$ is nonzero in the unbroken due to finite ultraviolet corrections.

At smaller values of $\lambda$ the expectation value of the squared scalar field experiences a jump that indicates the presence of a first-order phase transition. Therefore, we conclude that the point $(\beta_{\mathrm{latt}},\lambda,\kappa) = (4,10,5.215)$ is rather close to the end point of the phase-transition line. 

Below we simulate the gauge theory at the hopping parameter $\kappa  = 5.2$ which corresponds to the unbroken phase in the very vicinity of the critical endpoint. We observed that as the magnetic field becomes stronger, the vacuum of the theory shifts towards a deeper unbroken phase. Thus we use the magnetic field as a control parameter: by decreasing the magnetic field we gradually approach the conformal point from the side of the unbroken, ``scalar QED'' phase. This procedure allows us to make a smooth extrapolation to the nearly conformal point. Notice that by the construction, the magnetic field is set to vanish in the conformal point, so that there is no violation of the classical conformal symmetry which could otherwise emerge due to the presence of the background magnetic field.

\section{Electric current}
\label{sec:current}

The density of the electric current is given by the variation of the action~\eq{eq:S:AHM} with respect to the gauge field $\theta_{x\mu}$:
\beqn
j_{x\mu} = - 2 \avr{{\mathrm{Im}} \left[ \phi^*_x e^{i {(\theta_{x\mu} + \theta^B_{x\mu})}} \phi_{x+\hat\mu} \right]}\,.
\label{eq:j}
\eeqn

\begin{figure}[!thb]
\begin{center}
\includegraphics[scale=0.55,clip=true]{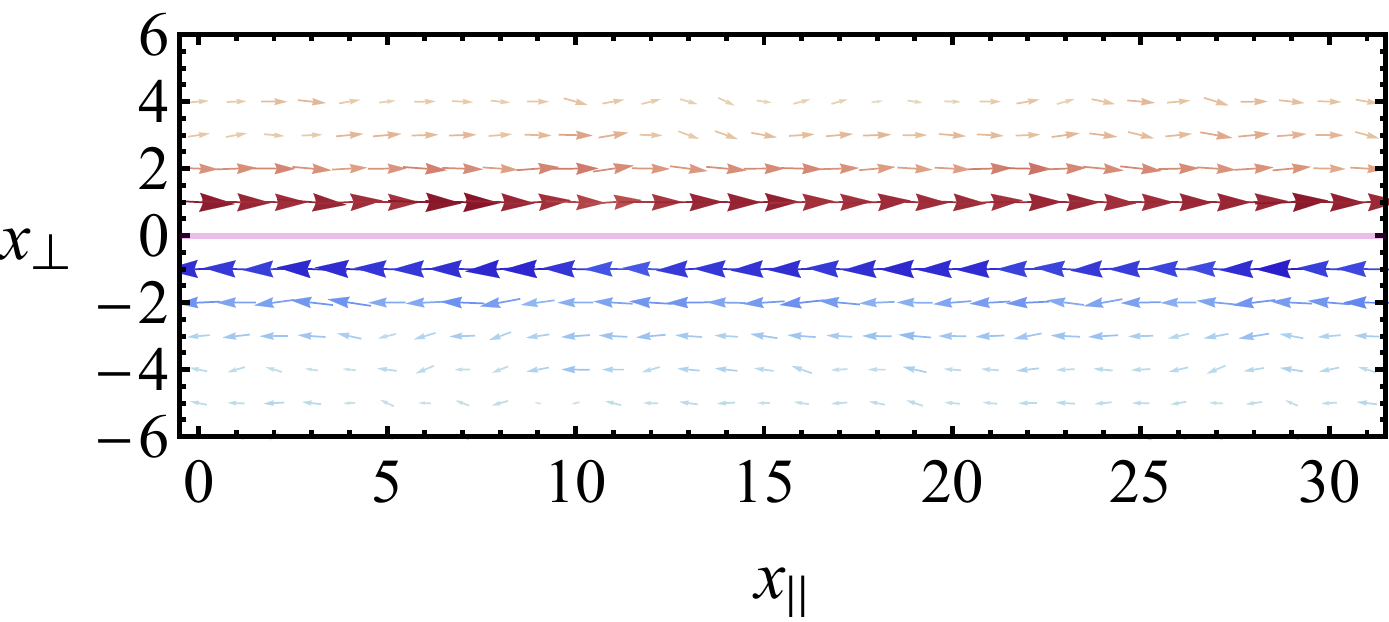}
\end{center}
\vskip -3mm
\caption{The CMEE: the electric current is generated in the vicinity of the Dirichlet boundary~\eq{eq:Dirichlet:boundary} shown by the pink horizontal line in the normal-tangential ($x_\perp, x_\|$) coordinate plane. The magnetic field $eB = 0.025$ is directed downwards with respect to the plane.}
\label{fig:current}
\end{figure}

We calculate numerically the electric current~\eq{eq:j} in the vicinity of the Dirichlet boundary. An example of the current density in the normal-tangential ($x_\perp, x_\|$) plane in the magnetic-field background is shown in Fig.~\ref{fig:current}. The magnetic field, which is directed downwards with respect to the plane, creates an electric current parallel to the boundary. In agreement with Eq.~\eq{eq:j:conformal}, the current is streamed in opposite directions at different sides of the boundary plane. The current density takes its maximum at the boundary and disappears in bulk, in a qualitative agreement with the space-dependent anomalous factor~\eq{eq:nu}.

\begin{figure}[!thb]
\begin{center}
\includegraphics[scale=0.5,clip=true]{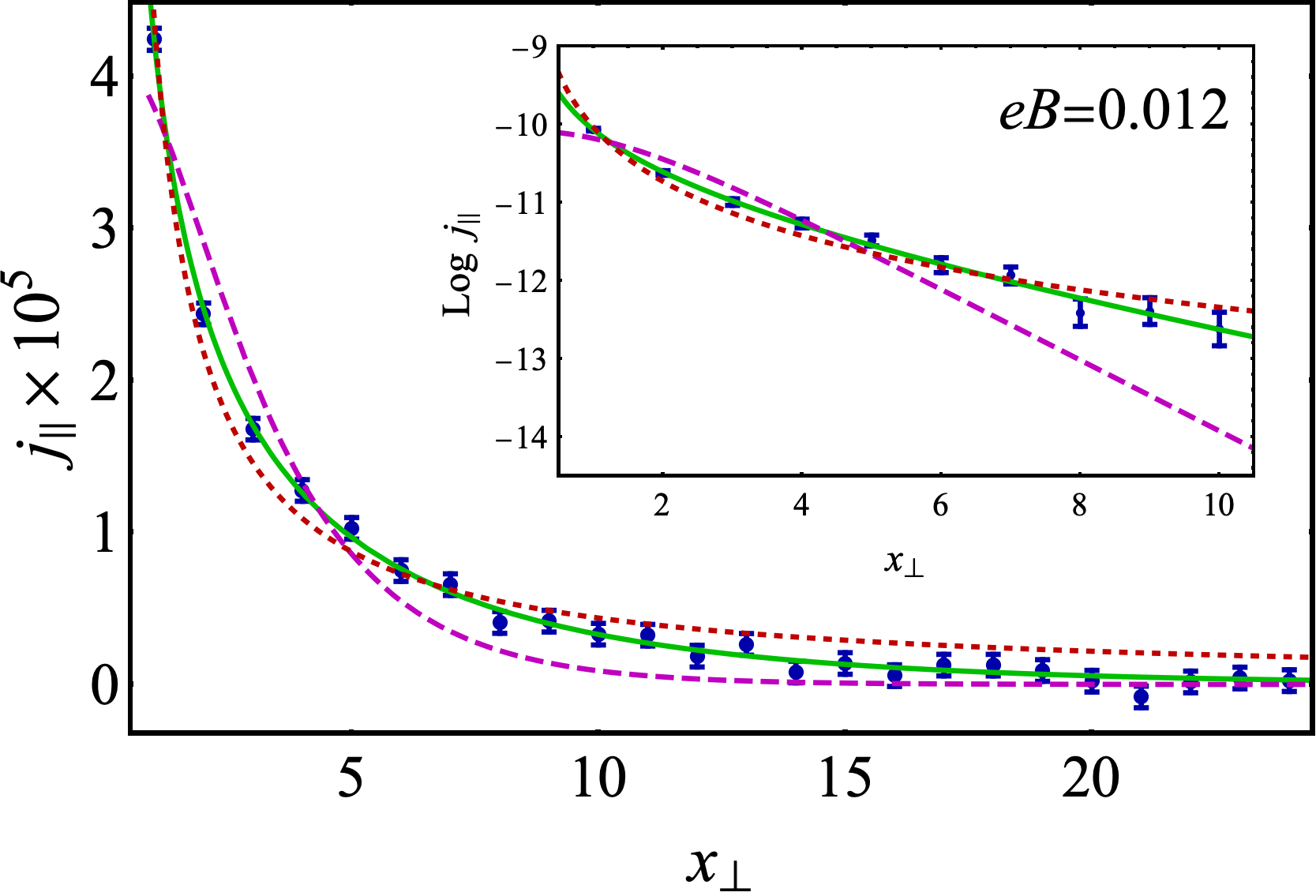}\\
\hskip 6mm (a) \\[5mm]
\includegraphics[scale=0.5,clip=true]{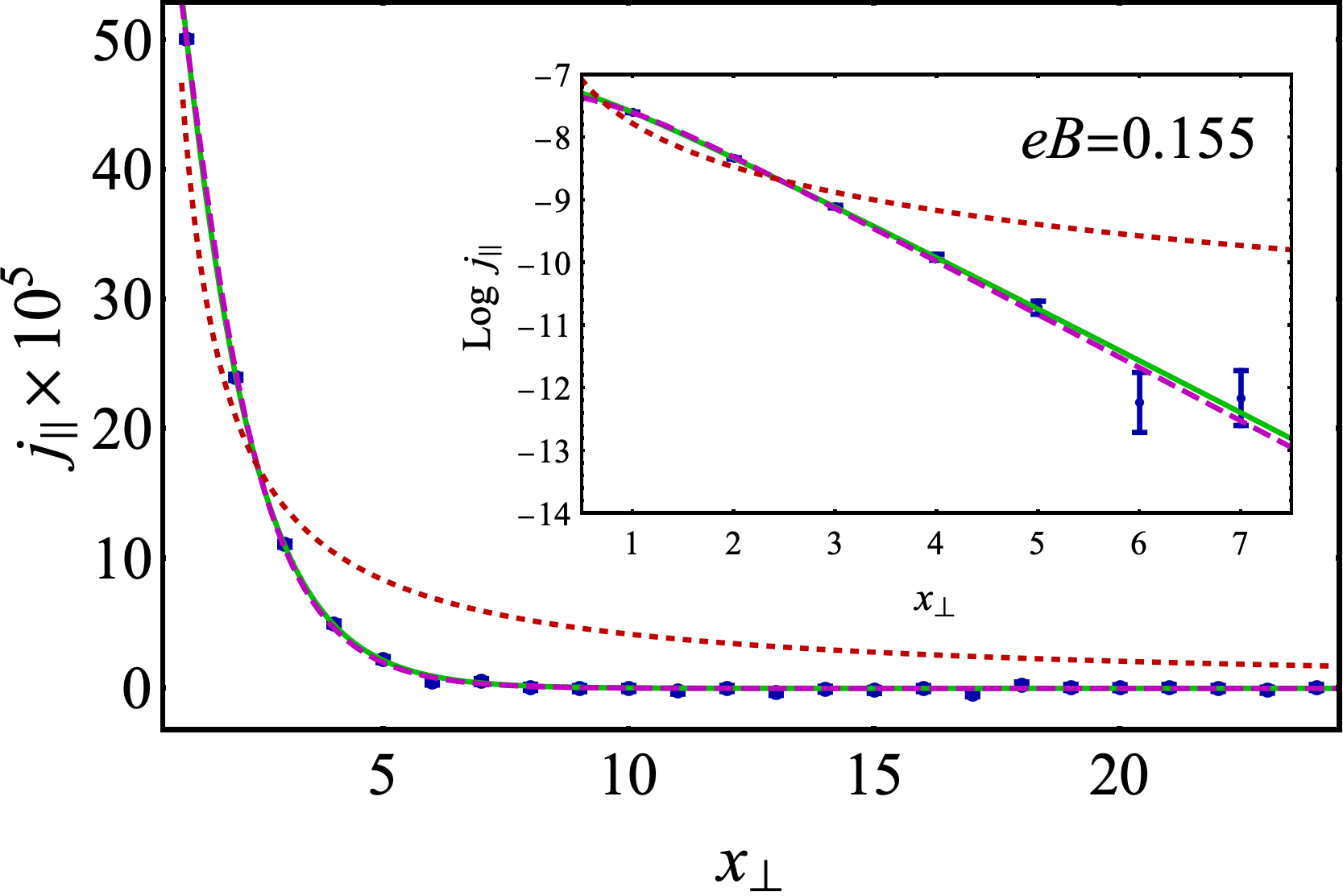}\\
\hskip 7mm (b)
\end{center}
\vskip -3mm
\caption{The tangential current density $j_\|$ as the function of the normal distance to the boundary $x_\perp$ (a) in the vicinity to the conformal point at the background magnetic field $eB = 0.012$ and (b) in a non-conformal region at $eB = 0.155$. The solid green, long-dashed magenta and short-dashed red lines are the best fits by Eqs.~\eq{eq:j:fit1}, \eq{eq:j:fit2} and \eq{eq:j:fit3}, respectively. The insets show the same plots in a logarithmical scale.}
\label{fig:fits:k3}
\end{figure}

The example of the behavior of the tangential current density $j_\|$ as the function of the distance to the boundary $x_\perp$ is shown in Fig.~\ref{fig:fits:k3} both in a near-to-conformal point (at a weak magnetic field) and in a far-from-conformal point (at a strong magnetic field). In order to compare the profile of the generated current with the prediction coming from the conformal anomaly~\eq{eq:nu}, we used three types of the fitting functions (written in the dimensionless lattice units):
\beqn
j^{\mathrm{fit,(1)}}_\|  & = & \gamma x_\perp^\nu\, e^{- M x_\perp} \,, 
\label{eq:j:fit1}\\
j^{\mathrm{fit,(2)}}_\|  & = & \frac{a}{\cosh m x_\perp} \,, 
\label{eq:j:fit2}\\
j^{\mathrm{fit,(3)}}_\|  & = & b x_\perp^{-1} \,, 
\label{eq:j:fit3}
\eeqn
that feature either an infrared cutoff or an ultraviolet cutoffs, or the both. Here $\gamma$, $\nu$, $M$, $a$, $b$, are $m$ are the fitting parameters.

The numerical data indicate that the fitting function~\eq{eq:j:fit1} is the best function (with $\chi^2/d.o.f. \sim 1$) which captures all features of the current density close to the boundary. This fact is especially well visible in the logarithmic plots shown in the insets of Fig.~\ref{fig:fits:k3}. 

Thus, outside of the conformal limit, the electric current density~\eq{eq:j:fit1} is characterized an anomalous power law $j_\| \sim x^\nu$ near the edge of the system ($x_\perp \to 0$) and by an exponential suppression factor $j_\| \sim e^{-x_\perp M}$ far away from the edge ($x_\perp M \gg 1$). 

In the conformal limit we expect that fitting parameters should converge to the theoretically predicted values
\beqn
\nu^{\mathrm{th}}_{\mathrm{conf}} = - 1,
\qquad
\gamma_{\mathrm{conf}} = \frac{eB }{24 \pi^2}\,,
\qquad 
M^{\mathrm{th}}_{\mathrm{conf}} = 0,
\label{eq:theor}
\eeqn
which can be read off from Eqs.~\eq{eq:j:conformal} and \eq{eq:nu:cQED}.

\begin{figure}[!thb]
\begin{center}
\includegraphics[scale=0.5,clip=true]{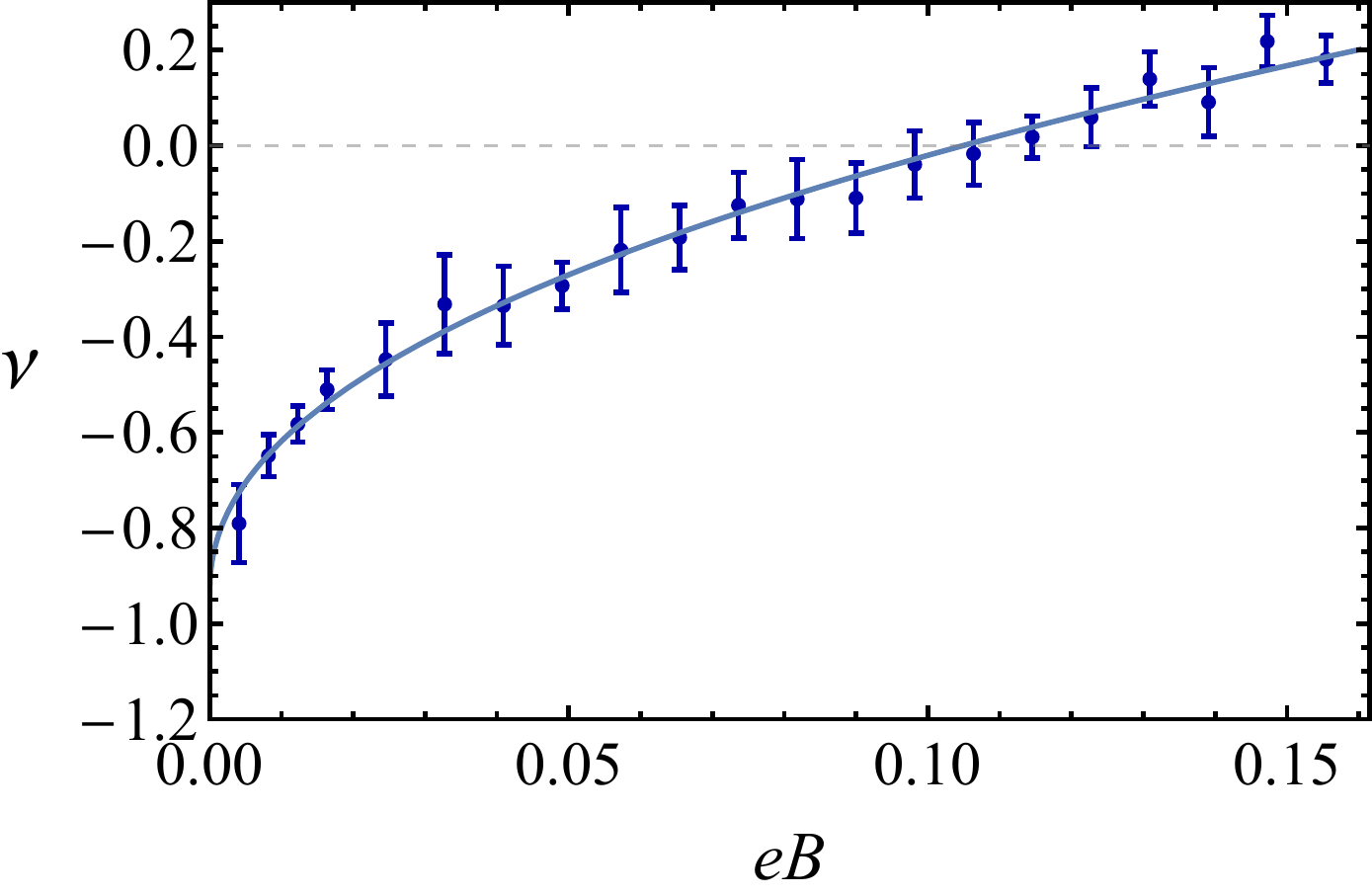} \\
\hskip 11mm (a) \\[5mm]
\includegraphics[scale=0.5,clip=true]{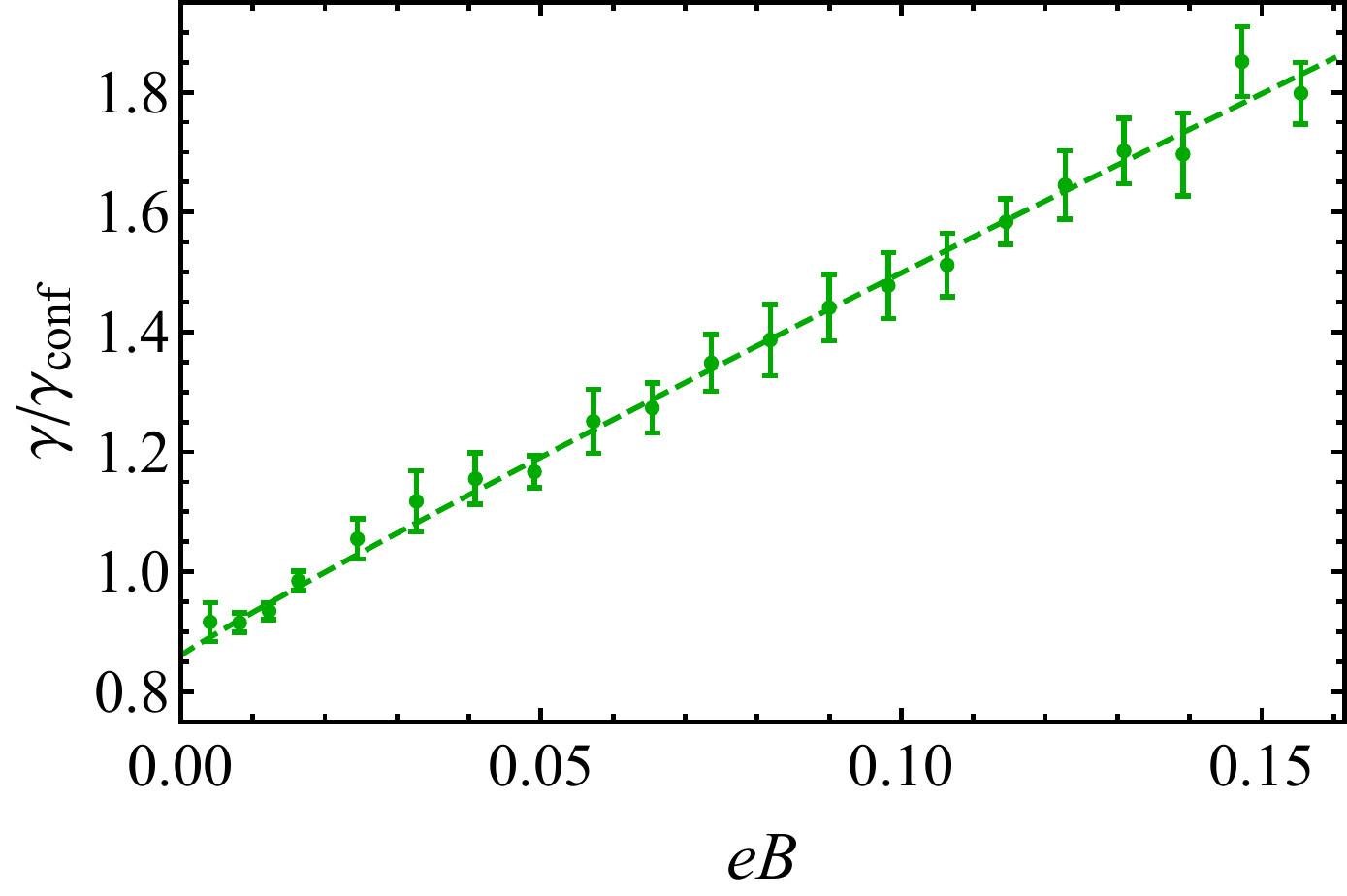} \\
\hskip 10mm (b) \\[5mm]
\includegraphics[scale=0.5,clip=true]{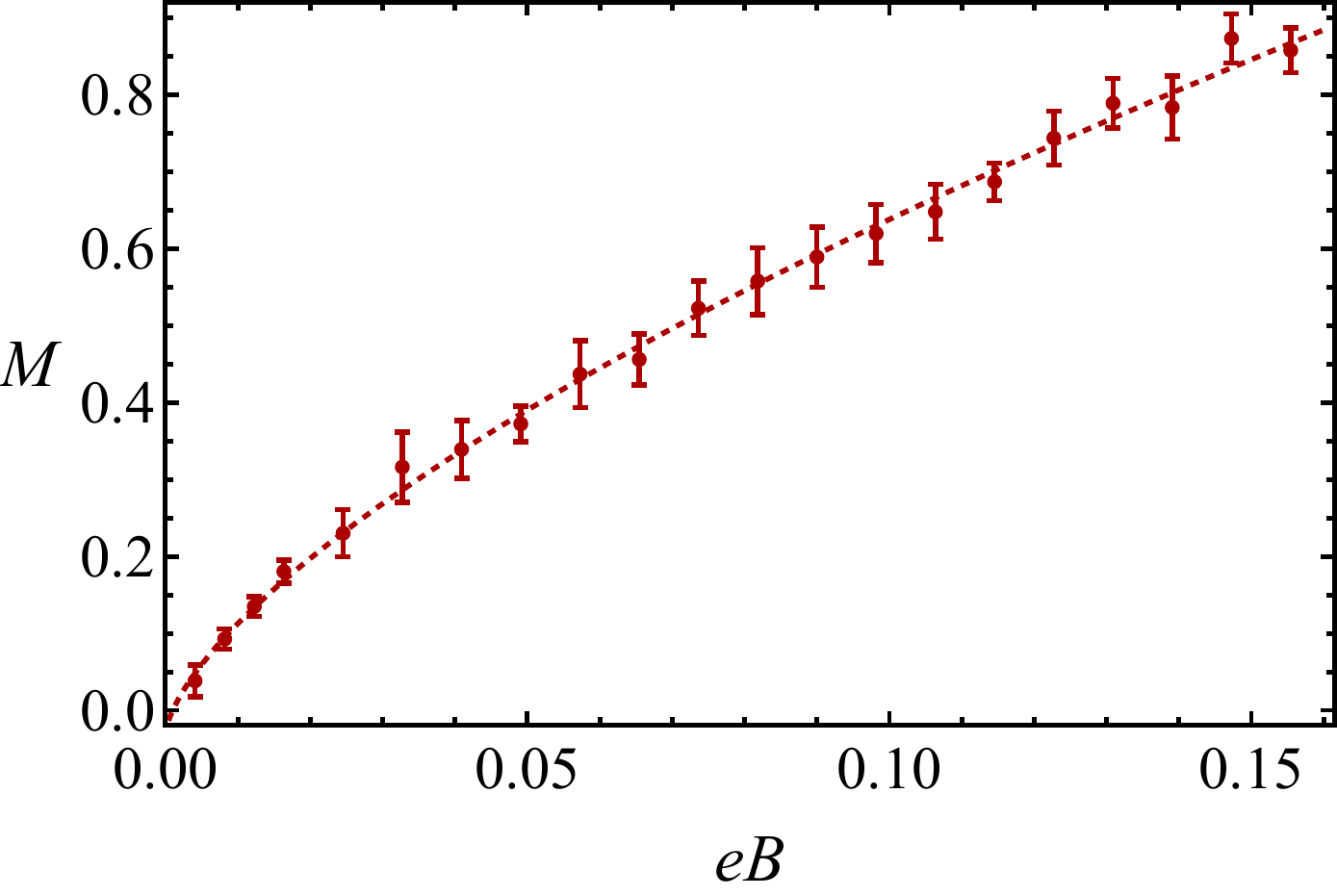} \\
\hskip 9mm (c)
\end{center}
\vskip -3mm
\caption{The best-fit parameters $\nu$, $\gamma$, $M$ of the density of the electric current~\eq{eq:j:fit1} as functions of the magnetic field strength~$B$.  The lines show the extrapolations~\eq{eq:fit:fit} to the conformal point $B = 0$.}
\label{fig:nu:eB}
\end{figure}

In Fig.~\ref{fig:nu:eB} we show the dependence of the best fit parameters $\nu$, $\gamma$ and $M$ of the current density~\eq{eq:j:fit1} as the function of the magnetic field $eB$. It is interesting to observe that all parameters are lively functions of magnetic field~$B$:
\begin{itemize}
\item[(i)] Figure~\ref{fig:nu:eB}(a) indicates that at small values of magnetic field the dimension of the power $\nu$ in the prefactor $x^\nu$ is negative in a qualitative agreement with the prediction~\eq{eq:nu} coming from the conformal anomaly. At certain magnetic field ($eB \simeq 0.11$) the power $\nu$ changes its sign, indicating that the boundary current gets its maximum not at the boundary $x_\perp = 0$, but rather at a finite distance $x^{\max}_\perp$ from the boundary. The analytical form of the current profile~\eq{eq:j:fit1} suggests that the maximal value of the current should be reached at the distance 
\beqn
x^{\max}_\perp = 
\left\{ 
\begin{array}{ll}
0, & \quad \nu \leqslant 0,\\
\frac{\nu}{M}, & \quad \nu > 0,
\end{array}
\right.
\label{eq:x:max}
\eeqn
where the power $\nu = \nu(B)$ and the mass $M = M(B)$ are both functions of the magnetic field $B$ according to Fig.~\ref{fig:nu:eB}. We cannot check this statement numerically since at our parameters the distance~\eq{eq:x:max} is still smaller than one lattice spacing.

\item[(ii)]  Figure~\ref{fig:nu:eB}(b) shows that the proportionality coefficient $\gamma$ is a linear function of magnetic field: this coefficient becomes larger as we move away from the conformal point.

\item[(iii)] Figure~\ref{fig:nu:eB}(c) illustrates that the mass $M$ tends to zero as the magnetic field decreases. The mass becomes large far away from the conformal point at high magnetic field. The quantity $M$ plays a role of the screening mass, which controls how fast the current diminishes at large distances from the physical boundary of the system.
\end{itemize}

The behavior of the mass parameter $M = M(B)$ is in agreement with our previous conclusion that the magnetic field may be used as a fine-tuning parameter which serves as a measure of the distance from the conformal point (given by the $B = 0$ point) in the parameter space. In order to extrapolate our results to the conformal point $B = 0$ we use the following fitting functions of the parameters $\cO = \nu, \gamma, M$ as the functions of the magnetic field $B$:
\beqn
\cO = \cO_{\mathrm{conf}} + \xi_\cO (eB)^{\alpha_\cO}.
\label{eq:fit:fit}
\eeqn
The corresponding best fits are shown in Fig.~\eq{fig:nu:eB} by the lines.  

Here $\cO_{\mathrm{conf}}$ are the values of the parameters extrapolated to the conformal limit. Our numerical calculations indicate that 
\beqn
\nu_{\mathrm{conf}} = - 0.94(8), \nonumber \\
M_{\mathrm{conf}} = -0.03(3), \\
\gamma_{\mathrm{conf}}/\gamma_{\mathrm{conf}}^{\mathrm{th}} = 0.86(2).
\nonumber
\eeqn
It is remarkable that the value of the power $\nu$ and the mass $M$ coincide, within the error bars, with the theoretical prediction coming from the conformal anomaly~\eq{eq:theor}. The proportionality coefficient $\gamma$ is within $15\%$ of the theoretical expectation. This small deviation may be caused by ultraviolet lattice artifacts which are not addressed in details in our exploratory study.

The slopes $\xi_\cO$ of the fitting functions~\eq{eq:fit:fit} are as follows:
\beqn
\xi_\nu = 2.6(2),
\quad
\xi_\gamma = 5.6(5),
\quad
\xi_M = 3.1(2).
\eeqn
We obtain the following values of the critical exponents $\alpha_\cO$,
\beqn
\alpha_\nu & = &  0.45(6) \simeq 1/2, \nonumber \\
\alpha_M & = & 0.67(3) \simeq 2/3, \\
\alpha_\gamma & = & 0.95(5) \simeq 1. \nonumber
\eeqn

According to Eq.~\eq{eq:j:fit1} the total current~\eq{eq:int:j} may be expressed via the Euler gamma function:
$J^{\mathrm{tot}}_\| = \gamma M^{-\nu-1} \Gamma(1+\nu)$.
Using this formula, as well as the numerical data for the parameters $\gamma$, $\nu$ and $M$ we calculate the total electric current flowing tangentially to the boundary.

\begin{figure}[!thb]
\begin{center}
\includegraphics[scale=0.55,clip=true]{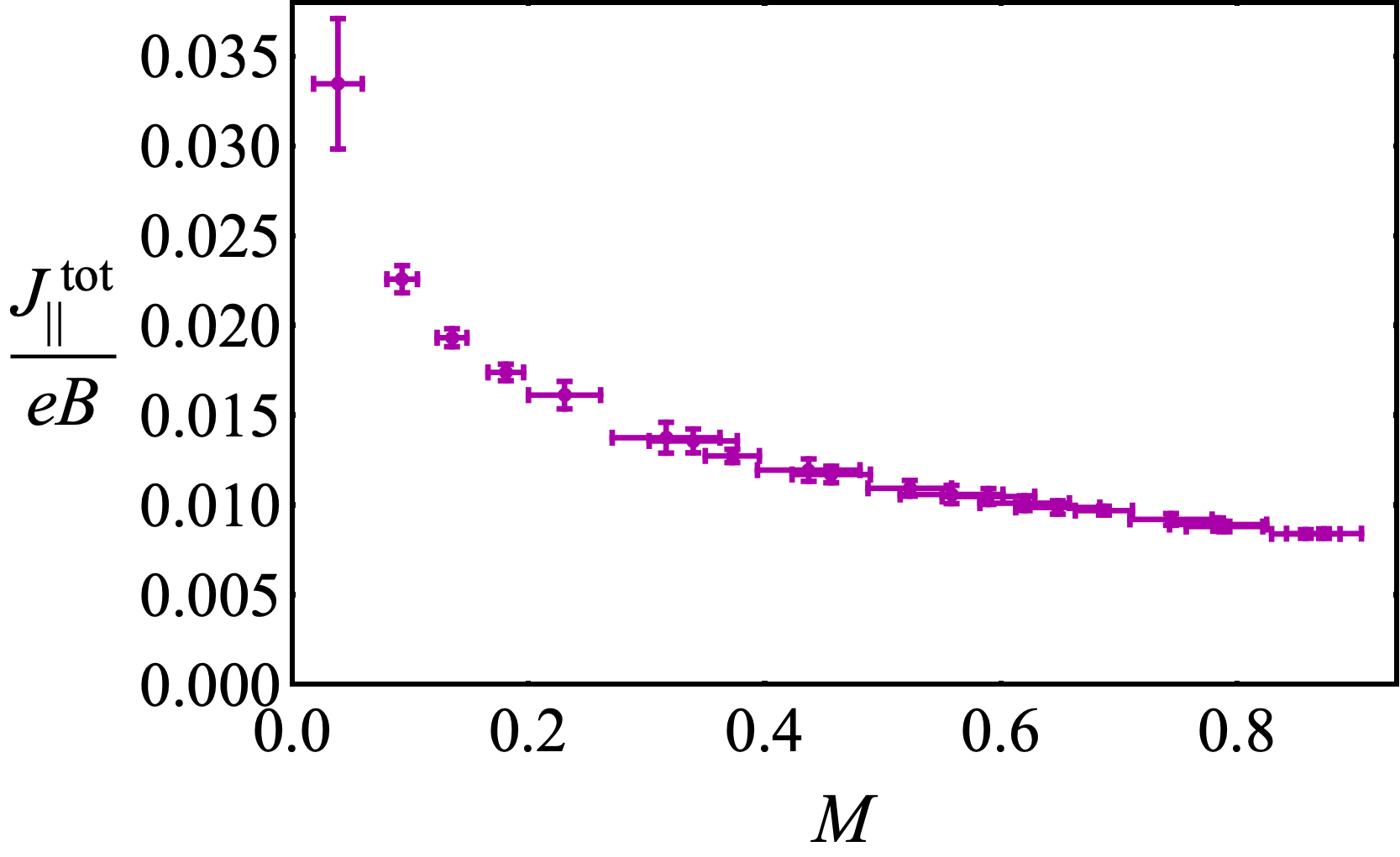} 
\end{center}
\vskip -3mm
\caption{The total current~\eq{eq:int:j} generated near the Dirichlet boundary by the background magnetic field $B$ as the function of the screening mass $M$. The total current is normalized by the magnetic field. The statistical errors are represented both by the vertical and horizontal bars reflecting the uncertainties in the current $J^{\mathrm{tot}}_\|$ and in the screening mass $M$, respectively.}
\label{fig:J:tot:int}
\end{figure}

The total electric current, normalized by the value of the magnetic field, is shown in Fig.~\ref{fig:J:tot:int}. First of all, we notice that at finite values of the screening mass $M$ the total generated current $J^{\mathrm{tot}}_\|$ is a finite quantity. The current grows rapidly as we approach the conformal limit, $M \to 0$. Far away from the conformal limit the total current is a slowly diminishing function of the increasing screening mass $M$.

\section{Conclusions}

In our paper we numerically demonstrate the existence of the Conformal magnetic edge effect (CMEE) which generates in a cold vacuum, in a presence of a background magnetic field, an anomalous electric current along the boundary (edge) of a geometrically bounded physical system~\eq{eq:j:conformal}. Similarly to the scale electromagnetic effects in a background gravitational field~\cite{Chernodub:2016lbo}, the strength of the boundary current is proportional to the beta function of the theory. The current emerges as a result of the conformal anomaly.

We carried out first-principle numerical simulations in the zero-temperature scalar QED with one species of the charged scalar field. Since this current emerges due to the conformal anomaly, we study its properties near a conformal point where the conformal symmetry is classically unbroken. To this end we determine the position of the conformal point in the coupling space of the model. Then we calculate the density of the electric current in the vicinity of this point and, finally, make an extrapolation to the conformal limit. 

In the conformal limit our results demonstrate a remarkable agreement, within an acceptable accuracy, with analytical calculations of Refs.~\cite{McAvity:1990we,Chu:2018ksb} given by Eq.~\eq{eq:j:conformal}. Outside the conformal limit, the current density close the edge of the system is described by a non-integer power law while far away from the edge the current density is exponentially suppressed~\eq{eq:j:fit1}. The total current, generated at a fixed external magnetic field, increases rapidly as one approaches the conformal point. 

\section*{Acknowledgments}

We thank Chong-Sun Chu, Claudio Corian\`o, Karl Landstei\-ner, Matteo Maria Maglio, Rong-Xin Miao, and Mar\'ia Vozmediano for interesting discussions. The research was carried out within the state assignment of the Ministry of Science and Education of Russia (Grant No. 3.6261.2017/8.9). The numerical simulations were performed at the computing cluster Vostok-1 of Far Eastern Federal University at Vladivostok, Russia.


\begin{thebibliography}{99}

\bibitem{Fukushima:2008xe} 
  K.~Fukushima, D.~E.~Kharzeev and H.~J.~Warringa,
``The Chiral Magnetic Effect,''
  Phys.\ Rev.\ D {\bf 78}, 074033 (2008)
  [arXiv:0808.3382 [hep-ph]].

\bibitem{Son:2004tq} 
  D.~T.~Son and A.~R.~Zhitnitsky,
  ``Quantum anomalies in dense matter,''
  Phys.\ Rev.\ D {\bf 70}, 074018 (2004)
  [hep-ph/0405216].

\bibitem{Landsteiner:2012kd} 
  K.~Landsteiner, E.~Megias and F.~Pena-Benitez,
 ``Anomalous Transport from Kubo Formulae,''
  Lect.\ Notes Phys.\  {\bf 871}, 433 (2013)
  [arXiv:1207.5808 [hep-th]].

\bibitem{Kharzeev:2013ffa} 
  D.~E.~Kharzeev,
  ``The Chiral Magnetic Effect and Anomaly-Induced Transport,''
  Prog.Part.\ Nucl.\ Phys.\  {\bf 75}, 133 (2014)
  [arXiv:1312.3348].

\bibitem{Chernodub:2016lbo} 
  M.~N.~Chernodub,
  ``Anomalous Transport Due to the Conformal Anomaly,''
  Phys.\ Rev.\ Lett.\  {\bf 117}, 141601 (2016).

\bibitem{Chernodub:2017jcp} 
  M.~N.~Chernodub, A.~Cortijo and M.~A.~H.~Vozmediano,
  ``Generation of a Nernst Current from the Conformal Anomaly in Dirac and Weyl Semimetals,''
  Phys.\ Rev.\ Lett.\  {\bf 120}, 206601 (2018)
  [arXiv:1712.05386 [cond-mat.str-el]].

\bibitem{ref:TJJ}
  R.~J.~Riegert,
  ``A Nonlocal Action for the Trace Anomaly,''
  Phys.\ Lett.\  {\bf 134B}, 56 (1984);
  E.~Mottola and R.~Vaulin,
 ``Macroscopic Effects of the Quantum Trace Anomaly,''
  Phys.\ Rev.\ D {\bf 74}, 064004 (2006)
  [gr-qc/0604051];
  R.~Armillis, C.~Corian\`o and L.~Delle Rose,
  ``Conformal Anomalies and the Gravitational Effective Action: The TJJ Correlator for a Dirac Fermion,''
  Phys.\ Rev.\ D {\bf 81}, 085001 (2010)
  [arXiv:0910.3381 [hep-ph]];
  C.~Corian\`o and M.~M.~Maglio,
  ``Exact Correlators from Conformal Ward Identities in Momentum Space and the Perturbative $TJJ$ Vertex,''
  arXiv:1802.07675 [hep-th].

\bibitem{McAvity:1990we} 
  D.~M.~McAvity and H.~Osborn,
  ``A DeWitt expansion of the heat kernel for manifolds with a boundary,''
  Class.\ Quant.\ Grav.\  {\bf 8}, 603 (1991).

\bibitem{Chu:2018ksb} 
  C.~S.~Chu and R.~X.~Miao,
  ``Anomaly Induced Transport in Boundary Quantum Field Theories,''
  arXiv:1803.03068 [hep-th];
  ``Anomalous Transport in Holographic Boundary Conformal Field Theories,''
  JHEP {\bf 07}, 005 (2018) [arXiv:1804.01648 [hep-th]].

\bibitem{ref:AHM}
  E.~H.~Fradkin and S.~H.~Shenker,
  ``Phase Diagrams of Lattice Gauge Theories with Higgs Fields,''
  Phys.\ Rev.\ D {\bf 19}, 3682 (1979);
   D.~Espriu and J.~F.~Wheater,
  ``On Spontaneous Symmetry Breaking in the Lattice Abelian Higgs Model,''
  Nucl.\ Phys.\ B {\bf 258}, 101 (1985);
  A.~Cruz, D.~Iniguez, L.~A.~Fernandez, A.~Munoz-Sudupe and A.~Tarancon,
  ``Study of the Coulomb-Higgs transition in the Abelian Higgs model,''
  Phys.\ Lett.\ B {\bf 416}, 163 (1998)
  [hep-lat/9708011].

\bibitem{ref:Gattringer} 
C. Gattringer, C.B. Lang, ``Quantum Chromodynamics on the Lattice'' (Springer-Verlag, Berlin Heidelberg, 2010).

\bibitem{ref:Omelyan}
I. P. Omelyan, I. M. Mryglod, and R. Folk, 
``Optimized Verlet-like algorithms for molecular dynamics simulations'', 
Phys. Rev. E {\bf 65}, 056706 (2002);
``Symplectic analytically integrable decomposition algorithms: classification, derivation, and application to molecular dynamics, quantum and celestial mechanics simulations'', 
Comput. Phys. Commun. {\bf 151}, 272 (2003).

\end{thebibliography}
\end{document}